\documentclass{PoS}

\usepackage{epsfig,multicol,multirow}
\usepackage{amsmath,subfigure,latexsym,amssymb}

\usepackage{fontenc,times,mathptmx,graphicx}

\newcommand {\beq} {\begin{equation}}
\newcommand {\eeq} {\end{equation}}
\newcommand {\beqa}{\begin{eqnarray}}
\newcommand {\eeqa}{\end{eqnarray}}

\newcommand {\tr}{{\rm tr\,}}

\title{Supersymmetry non-renormalization theorem from a computer 
and the AdS/CFT correspondence}

\ShortTitle{SUSY non-renormalization theorem from a computer and the AdS/CFT}

\author{\speaker{Masazumi Honda}$^{a}$, Goro Ishiki$^{b,c}$, 
Sang-Woo Kim$^{b,c,d}$,
Jun Nishimura$^{a,b}$ \hspace{3cm} and Asato Tsuchiya$^{e}$
\vspace*{0.5cm} \\
\llap{$^a$}Department of Particle and Nuclear Physics,\\
Graduate University for Advanced Studies (SOKENDAI),\\
Tsukuba, Ibaraki 305-0801, Japan\\
\llap{$^b$}KEK Theory Center, 
High Energy Accelerator Research Organization (KEK),\\
Tsukuba, Ibaraki 305-0801, Japan\\
\llap{$^c$}Center for Quantum Spacetime (CQUeST),\\
Sogang University, Seoul 121-742, Korea\\
\llap{$^d$}Department of Physics, Osaka University,\\
Toyonaka, Osaka 560-0043, Japan\\
\llap{$^e$}Department of Physics, Shizuoka University,\\
836 Ohya, Suruga-ku, Shizuoka 422-8529, Japan
\vspace*{0.5cm} \\
\email{mhonda@post.kek.jp},
\email{ishiki@post.kek.jp},
\email{sang@het.phys.sci.osaka-u.ac.jp},\\
\email{jnishi@post.kek.jp},
\email{satsuch@ipc.shizuoka.ac.jp}}

\abstract{We perform Monte Carlo calculation
of correlation functions in 4d $\mathcal{N}=4$
super Yang-Mills theory on $R\times S^3$ in the planar limit.
In order to circumvent the well-known problem of lattice SUSY,
we adopt the idea of a novel large-$N$ reduction,
which reduces the calculation to that of corresponding 
correlation functions in the plane-wave matrix model
or the BMN matrix model.
This model is a 1d gauge theory with 16 supersymmetries,
which can be simulated in a manner similar to 
the recent studies of the D0-brane system.
We study two-point and three-point functions of
chiral primary operators at various coupling constant,
and find that they agree with the free theory results up to 
overall constant factors.
The ratio of the overall factors for 
two-point and three-point functions agrees with
the prediction of the AdS/CFT correspondence.
}

\FullConference{The XXVIII International Symposium on Lattice Field Theory, 
Lattice2010\\
June 14-19, 2010\\
Villasimius, Italy}

\begin{document}

\section{Introduction}
The gauge-gravity duality \cite{AdS-CFT} has been one of the
most important subjects in string theory over the past decade.
The most typical example is
the so-called AdS/CFT correspondence
between type IIB superstring theory on $AdS_5\times S^5$
and 4d ${\cal N}=4$ U($N$) super Yang-Mills theory (SYM).
Even in this case, however,
a complete proof of the duality is still missing.
In particular, the region described by the classical supergravity 
on the string theory side
corresponds to the strongly coupled region in the planar large-$N$ limit 
on the SYM side.
In order to study the strongly coupled 4d ${\cal N}=4$ SYM 
from first principles, 
one needs to have a non-perturbative formulation such as the lattice QCD.
The problem here is that the lattice regularization necessarily breaks 
translational symmetry, which is included in the supersymmetry (SUSY).
In order to restore SUSY in the continuum limit, 
one generally has to fine-tune parameters in the lattice action.
In fact any lattice formulations of 4d ${\cal N}=4$ SYM proposed so far
seem to require fine-tuning of at least 
three parameters \cite{latticeSUSY_N4}.

Since 4d ${\cal N}=4$ SYM has conformal symmetry,
the theory on $R^4$ is equivalent to the theory on $R\times S^3$ 
through conformal mapping.
The novel large-$N$ reduction \cite{Ishii:2008ib} 
connects the planar large-$N$ limit of this theory to a reduced model,
which can be obtained by shrinking the $S^3$ to a point.
The resulting one-dimensional gauge theory with 16 supercharges can
be studied by using the Fourier-mode simulation \cite{Hanada:2007ti} 
as in recent studies of the
D0-brane system \cite{BFSS_sim}.
Thus we can perform Monte Carlo calculations
in 4d ${\cal N}=4$ SYM respecting SUSY maximally and 
without fine-tuning.\footnote{See refs.\ \cite{Hanada:2010kt}
for proposals for finite $N$.
}

In this article 
we present explicit results for correlation functions 
of chiral primary operators (CPOs) in 
4d $\mathcal{N}=4$ SYM.\footnote{See
ref.\ \cite{Nishimura:2009xm} for some preliminary results on 
the Wilson loop.}
In particular, we find that the two-point and three-point functions
agree with the free theory results up to overall constant factors
even at fairly strong coupling.
Moreover the ratio of the overall factors agrees with the prediction
of the AdS/CFT correspondence.\footnote{There are 
also Monte Carlo studies of the 4d ${\cal N}=4$ SYM based on 
matrix quantum mechanics of 6 bosonic commuting matrices
\cite{Berenstein_sim}, which give results consistent with
the AdS/CFT for the three-point functions of CPOs.}
 

\section{Large-$N$ reduction for ${\cal N}=4$ SYM on $R\times S^3$}
\label{sec:large-N-red}
Let us first discuss the novel large-$N$ reduction
for ${\cal N}=4$ SYM on $R\times S^3$.
By collapsing the $S^3$ to a point,
we obtain the plane wave matrix model (PWMM) 
or the BMN matrix model \cite{Berenstein:2002jq}\footnote{Properties 
of this model at finite temperature are studied at weak coupling
\cite{Kawahara:2006hs,eff_PWMM}
and at strong coupling \cite{Catterall:2010gf}.},
whose action is given by
\begin{eqnarray}
S_{\rm PW}
&=& \frac{1}{g_{\rm PW}^2}
\int
dt \, \tr 
\left[\frac{1}{2}(D_tX_M)^2-\frac{1}{4}[X_M,X_N]^2
+\frac{1}{2}\Psi^{\dagger} D_t \Psi
-\frac{1}{2}\Psi^{\dagger}\gamma_M[X_M,\Psi] \right.\nonumber\\
&~& \quad \quad \left.+\frac{\mu^2}{2}(X_i)^2
+\frac{\mu^2}{8}(X_a)^2 +i\mu\epsilon_{ijk}X_iX_jX_k
+i\frac{3\mu}{8}\Psi^{\dagger}\gamma_{123}\Psi \right] \ .
\label{pp-action}
\end{eqnarray}
Here the parameter $\mu$ is 
related to the radius of $S^3$ as $R_{S^3}=\frac{2}{\mu}$,
and the covariant derivative is defined by
$D_t=\partial_t-i[A, \ \cdot \ ]$,
where $A(t)$, as well as $X_M(t)$ and $\Psi (t)$, 
is an $N\times N$ hermitian matrix.
The range of indices is given by 
$1 \le M,N \le 9$, $1 \le i,j,k \le 3$ and $4 \le a \le 9$.
The model has the SU$(2|4)$ symmetry with 16 supercharges.

In fact the model possesses many discrete vacua 
representing multi fuzzy spheres, which are given explicitly by
\begin{equation}
X_i=\mu \bigoplus_{I=1}^{\nu}
\Bigl( L_i^{(n_I)}\otimes {\bf 1}_{k_I} \Bigr) \ 
\quad \mbox{with}\ \ \sum_{I=1}^{\nu}n_Ik_I=N \ ,
 \label{background}
\end{equation}
where $L_i^{(r)}$ are
the $r$-dimensional irreducible representation of the SU$(2)$ algebra 
$[L_i^{(r)},L_j^{(r)}]=i \, \epsilon_{ijk} \, L_k^{(r)}$.
These vacua preserve the SU$(2|4)$ symmetry, and are all degenerate.

In order to retrieve the planar ${\cal N}=4$ SYM on $R \times S^3$,
one has to pick up a particular background from (\ref{background}),
and consider the theory (\ref{pp-action}) around it.
Let us consider the case
\begin{equation}
k_I=k \ , \quad n_I=n+I-\frac{\nu+1}{2} \quad \quad
\mbox{for\ \ $I =1, \cdots , \nu$} \ ,
\label{our background}
\end{equation}
and take the large-$N$ limit in such a way that
\begin{eqnarray}
k\rightarrow \infty \ , \
\frac{n}{\nu} \rightarrow \infty \ , \ 
\nu\rightarrow\infty \ ,
\quad \mbox{with} \;\;
\lambda_{\rm PW} \equiv \frac{g_{\rm PW}^2 k}{n}
\; \; \mbox{fixed} \ .
\label{limit}
\end{eqnarray}
%
Then the resulting theory is claimed
\cite{Ishii:2008ib} to be equivalent\footnote{
See refs.\ \cite{earlier_novel} for earlier studies that led to 
this proposal. 
This equivalence was 
checked at finite temperature in the weak coupling regime \cite{eff_PWMM}.
It has also been extended to general group manifolds and 
coset spaces \cite{group-coset}.}
to the planar limit of ${\cal N}=4$ SYM on $R \times S^3$
with the 't Hooft coupling constant given by
\beq
\lambda_{\rm SYM} =2\pi^2 \lambda_{\rm PW}(R_{S^3})^3 
=\frac{16 \pi^2 k }{n}
\frac{g_{\rm PW}^2}{\mu^3} \ .
\label{lambda-sym}
\eeq
%

The above equivalence may be viewed as
an extension of the large-$N$ reduction \cite{EK},
which asserts that the large-$N$ gauge theories can be 
studied by dimensionally reduced models.
The original idea for theories compactified on a torus
can fail due to
the instability of the U(1)$^D$ symmetric vacuum
of the reduced model \cite{Bhanot}.
This problem is avoided in the novel proposal
since the PWMM is a massive theory and the vacuum preserves the maximal SUSY.
Since the planar limit is taken in the reduced model,
the instanton transition to other vacua
and the ``fuzziness'' of the spheres are suppressed.
Viewed as a regularization of the ${\cal N}=4$ SYM on $R\times S^3$,
the present formulation respects the maximal ${\rm SU}(2|4)$ symmetry 
(with 16 supercharges) of the PWMM, and 
in the limit (\ref{limit})
the symmetry is expected to enhance to the full 
superconformal ${\rm SU}(2,2|4)$ symmetry 
(with 32 supercharges).
Since any kind of UV regularization breaks the conformal symmetry,
this regularization is optimal from the viewpoint of preserving SUSY.

\section{Correlation functions in 4d $\mathcal{N}=4$ SYM}
\label{sec:corr-fun-sym}
As simple examples of $1/2$ BPS operators,
let us consider the CPOs given by
\begin{equation}
\mathcal{O}^{R^{4}}_{\Delta}(x)= 
T_{a_{1}\cdots a_{\Delta}}\ 
\tr\left( X^{R^{4}}_{a_{1}}X^{R^{4}}_{a_{2}}\cdots X^{R^{4}}_{a_{\Delta}}(x) 
\right) \ ,
\end{equation}
where $T_{a_{1}\cdots a_{\Delta}}$ is a symmetric traceless tensor
and $X_{a}^{R^{4}}$ represents 
the six scalars in 4d $\mathcal{N}=4$ SYM on $R^{4}$.
Thanks to the conformal symmetry, the forms of two-point and three-point 
functions of the CPOs are determined as
\begin{eqnarray}
\left\langle\mathcal{O}^{R^{4}}_{\Delta}(x_{1}) 
\mathcal{O}^{R^{4}}_{\Delta}(x_{2}) \right\rangle 
&=& c_{\Delta} \left\langle\mathcal{O}^{R^{4}}_{\Delta}(x_{1}) 
\mathcal{O}^{R^{4}}_{\Delta}(x_{2}) \right\rangle _{\rm free} \ , 
\nonumber \\
\left\langle\mathcal{O}^{R^{4}}_{\Delta_{1}}(x_{1}) 
\mathcal{O}^{R^{4}}_{\Delta_{2}}(x_{2}) 
\mathcal{O}^{R^{4}}_{\Delta_{3}}(x_{3}) \right\rangle 
&=& c_{\Delta_{1}\Delta_{2}\Delta_{3}} \left\langle
\mathcal{O}^{R^{4}}_{\Delta_{1}}(x_{1}) 
\mathcal{O}^{R^{4}}_{\Delta_{2}}(x_{2}) 
\mathcal{O}^{R^{4}}_{\Delta_{3}}(x_{3}) \right\rangle _{\rm free} \ , 
\label{O-Ofree}
\end{eqnarray}
where $c_{\Delta}$ and $c_{\Delta_{1}\Delta_{2}\Delta_{3}}$ are 
over-all constants depending on $\lambda_{\rm SYM}$ in general,
and $\langle\cdots \rangle _{\rm free}$ denotes the results of free theory.
The analysis on the gravity side suggests \cite{Lee:1998bxa}
\begin{equation}
\left. \frac{c_{\Delta_{1}\Delta_{2}\Delta_{3}}}
{\sqrt{c_{\Delta_{1}}c_{\Delta_{2}}c_{\Delta_{3}}}} 
\right|_{N\rightarrow\infty,\lambda_{\rm SYM}\rightarrow\infty}
=\left. \frac{c_{\Delta_{1}\Delta_{2}\Delta_{3}}}
{\sqrt{c_{\Delta_{1}}c_{\Delta_{2}}c_{\Delta_{3}}}} 
\right|_{N\rightarrow\infty,\lambda_{\rm SYM}\rightarrow 0}=1
\quad \mbox{for}\ \ ^{\forall}\Delta_{i} \ . 
\label{ratio_gravity}
\end{equation}

In order to relate the above operators
to those in the PWMM, 
we first perform the conformal mapping\footnote{ 
The metrics of $R^4$ and $R\times S^3$
are related as
$ds^2_{R^4}  = dr^2+r^2d\Omega_3^2  =e^{\mu t}ds^2_{R\times S^3}$, 
where $r=\frac{2}{\mu}e^{\frac{\mu}{2}t}$.
The transformation of the CPOs is given by
$\mathcal{O}^{R\times S^{3}}_{\Delta}=
e^{\frac{\Delta}{2}\mu t} \mathcal{O}^{R^{4}}_{\Delta} $.}
from $R^{4}$ to $R\times S^{3}$.
Then the $M$-point functions of the 
CPO $\mathcal{O}^{R\times S^{3}}_{\Delta_{i}}$ on $R\times S^{3}$ 
are related to those 
in PWMM as
\begin{equation}
\int\frac{d\Omega_{3}^{(1)}}{2\pi^{2}}\cdots 
\int\frac{d\Omega_{3}^{(M)}}{2\pi^{2}}
\left\langle \mathcal{O}_{\Delta_{1}}^{R\times S^{3}}
( t_{1},\Omega_{3}^{(1)}) \cdots 
\mathcal{O}_{\Delta_{M}}^{R\times S^{3}}
( t_{M},\Omega_{3}^{(M)} ) \right\rangle 
=\frac{1}{n^M\nu} \left\langle \mathcal{O}_{\Delta_1}^{\rm PW}(t_1)
\cdots \mathcal{O}_{\Delta_M}^{\rm PW}(t_M) \right\rangle  \ ,
\label{correspondence}
\end{equation}
where we have defined
$\mathcal{O}^{\rm PW}_{\Delta}(t)= T_{a_{1}\cdots a_{\Delta}}\ 
\tr\Big( X_{a_{1}}X_{a_{2}}\cdots X_{a_{\Delta}}(t) \Big)$ 
\cite{Ishii:2008ib}.

We calculate the two-point functions
$\Big\langle\tr Z^{2}(t_{1}) \, \tr Z^{\dag 2}(t_{2})\Big\rangle$,
where 
$Z=\frac{1}{\sqrt{2}}(X_{4}+iX_{5})$,
and the three-point functions
$\Big\langle \tr
\Big( X_{4}X_{5}(t_{1})\Big) \, \tr
\Big( X_{5}X_{6}(t_{2})\Big) \, \tr \Big( X_{6}X_{4}(t_{3}) \Big) 
\Big\rangle$.
The CPOs we consider here have $\Delta =2$, and
%
the AdS/CFT predicts
$c_{222}=c_{2}^{3/2}$,
which we test by Monte Carlo calculations.

\section{Monte Carlo method}
\label{sec:monte}
In order to simulate the PWMM (\ref{pp-action}),
we compactify the $t$-direction to a circle of circumference $\beta$.
Since we are interested in the properties at zero temperature,
we impose periodic boundary conditions 
on both scalars $X_{i}(t)$ and fermions $\Psi_{\alpha}(t)$,
which keep SUSY intact.
In Fourier-mode simulation \cite{Hanada:2007ti},
we first fix the gauge symmetry completely by choosing
$A(t) = \frac{1}{\beta}{\rm diag} (\alpha_{1},\cdots ,\alpha_{N})$
with $-\pi <\alpha_{a}\leq \pi$,
and then make a Fourier expansion 
\ $X_{i}(t) = \sum_{n=-\Lambda}^{\Lambda} 
\tilde{X}_{i,n}e^{i\omega nt}\ (\omega\equiv \frac{2\pi}{\beta})$ 
and similarly for the fermions.
The upper bound $\Lambda$ on the Fourier modes plays 
the role of the UV cutoff.
The original PWMM can be retrieved by just taking 
the limits $\beta\rightarrow\infty$
and $\frac{\Lambda}{\beta} \rightarrow \infty$
since there are neither UV nor IR divergences.
The model regularized by finite $\beta$ and $\Lambda$ 
can be simulated by the RHMC algorithm.
This method has been applied extensively
to the D0-brane system corresponding to $\mu=0$,
and the results confirmed the gauge/gravity duality 
for various observables \cite{BFSS_sim}.\footnote{See
refs.\ \cite{Catterall-Wiseman} for Monte Carlo calculations
based on the lattice regularization.
}

Since the parameter $g_{\rm PW}^{2}$ in the action (\ref{pp-action})
can be scaled away by appropriate redefinition of fields and parameters,
we take $g_{\rm PW}^{2}N=1$ without loss of generality
as 
in refs.\ \cite{BFSS_sim}.
In this convention one finds from eq.\ (\ref{lambda-sym}) that
the small (large) 
$\mu$ region in the PWMM corresponds to the strong (weak) coupling 
region in the 4d $\mathcal{N}=4$ SYM.

\begin{figure}[htbp]
  \begin{center}
   \includegraphics[width=7.2cm]{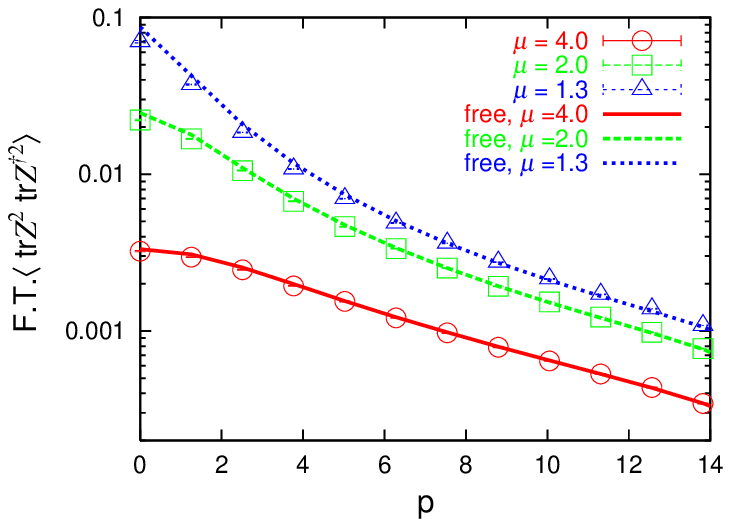}
   \includegraphics[width=7.2cm]{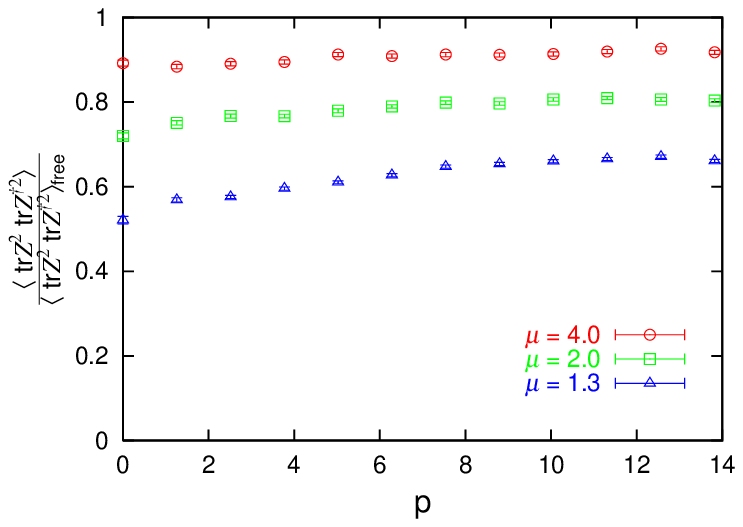}
  \end{center}
  \caption{(Left) The two-point function 
$\Big\langle \tr\widetilde{Z^{2}}(p)\ 
\tr \widetilde{Z^{\dag 2}}(-p) \Big\rangle $ 
is plotted 
in the log scale.
The curves represent 
the corresponding free theory results
multiplied by 0.919, 0.799, 0.647 for $\mu=4.0,2.0,1.3$,
respectively.
(Right) The ratio of the two-point function
to the corresponding free theory result is plotted 
in the linear scale
for $\mu=4.0,2.0,1.3$.
}
  \label{fig:ch2}
\end{figure}
\begin{figure}[htbp]
  \begin{center}
   \includegraphics[width=7.2cm]{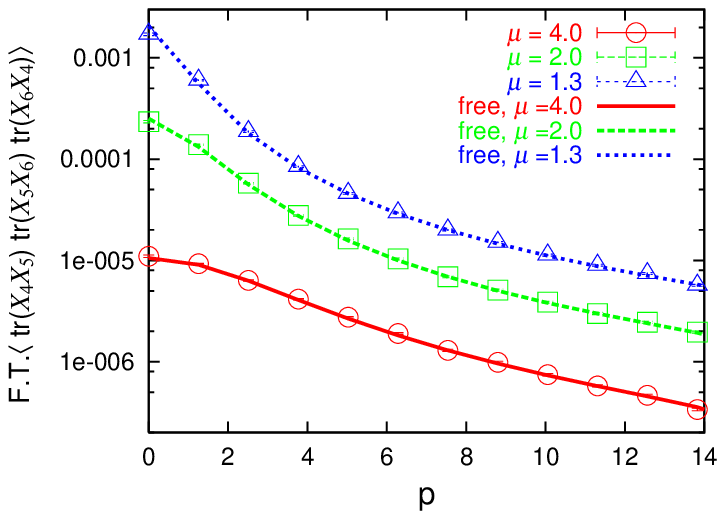}
   \includegraphics[width=7.2cm]{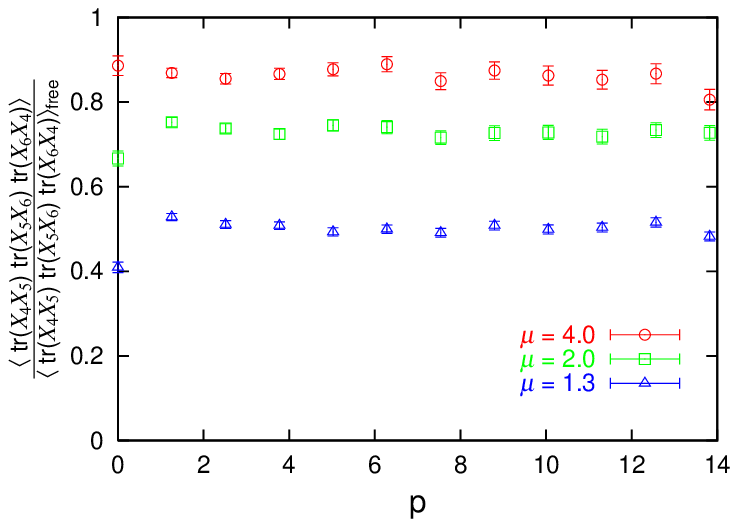}
  \end{center}
  \caption{(Left) The three-point function
$\Big\langle \tr\left( 
\widetilde{{X}_{4} X_{5}}(p)\right) \, 
\tr\left( \widetilde{ X_{5}X_{6} }(0)\right) \, 
\tr\left( \widetilde{X_{6}X_{4}}(-p)\right) \Big\rangle $
is plotted 
in the log scale. 
The curves represent 
the corresponding free theory results
multiplied by 0.850, 0.716, 0.491 for $\mu=4.0,2.0,1.3$,
respectively.
(Right) The ratio of the three-point function
to the corresponding free theory result is plotted 
in the linear scale
for $\mu=4.0,2.0,1.3$.}
  \label{fig:ch3}
\end{figure}

\section{Numerical results}
\label{sec:results}

The parameters describing the background (\ref{our background})
are chosen as $n=\frac{3}{2},\nu =2,k=2$, which corresponds 
to the matrix size $N=6$.
We use (\ref{background})
with (\ref{our background}) 
as the initial configuration
and check that no
transition to other vacua occurs during the simulation.
The values of $\mu$ we use are
$\mu=4.0,2.0,1.3$, which correspond 
to $\lambda_{\rm SYM}\simeq 0.55,4.39,16.0$, respectively,
in the chosen background.
Thus we cover a wide range of the coupling constant.
The regularization parameters in the $t$-direction are
taken as $\beta =5.0,\Lambda =12$ for all cases.

In fig.\ \ref{fig:ch2} (Left)
we plot the two-point function\footnote{The Fourier transform
of an operator $\mathcal{O}(t)$ is defined
as $\tilde{\mathcal{O}}(p)=\frac{1}{\beta} \int_{0}^{\beta}dt \, 
\mathcal{O}(t) \, e^{-ipt}$.}
$\Big\langle \tr \widetilde{Z^{2}}(p) \, 
\tr \widetilde{Z^{\dag 2}}(-p) \Big\rangle$.
We find that the results agree well --- up to overall constants 
depending on $\mu$ ---
with
the corresponding free theory results,
which are obtained by just switching off the interaction terms
in the reduced model with the same regularization parameters.
In fig.~\ref{fig:ch2} (Right)
we plot the ratio to the free theory results.
Figure \ref{fig:ch3} shows similar results 
for the three-point function  
$\Big\langle \tr\left( 
\widetilde{{X}_{4} X_{5}}(p)\right) \, 
\tr\left( \widetilde{ X_{5}X_{6} }(0)\right) \, 
\tr\left( \widetilde{X_{6}X_{4}}(-p)\right) \Big\rangle $.

\FIGURE{
\epsfig{file=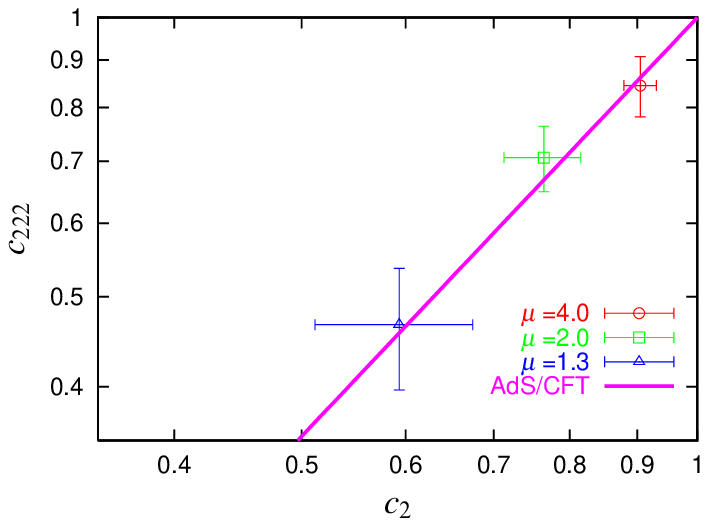,width=.45\textwidth}
\caption{The overall constants corresponding to 
$c_{2}$ and $c_{222}$ 
in eq.\ (\protect\ref{O-Ofree})
are plotted in the log-log scale.
The straight line presents the relation
$c_{222}=c_{2}^{3/2}$ predicted by the AdS/CFT.
}
\label{fig:ch23}
}

We can extract the the overall constants corresponding to
$c_{2}$ and $c_{222}$ in eq.\ (\ref{O-Ofree}) from
figs.~\ref{fig:ch2} and \ref{fig:ch3}, respectively.
Since the data on the right panels are not completely constant
in $p$, we take the maximum and minimum values
as the upper and lower bounds of the estimates.
In fig.\ \ref{fig:ch23}
we plot the overall constants 
obtained in this way for three values of $\mu$.
The data points represent the mean value of 
the upper and lower bounds.
We find that our results 
for various coupling constants
lie on the straight line which represents
the prediction $c_{222}=c_{2}^{3/2}$
from the AdS/CFT.
Our results therefore suggest that the relation
(\ref{ratio_gravity}) holds also at intermediate 
coupling constants.


\section{Summary and discussions}

We have made the first attempt to investigate
nonperturbative properties of the 4d $\mathcal{N}=4$ SYM
from first-principle calculations.
Our formulation is considered optimal in preserving SUSY,
which seems to give us the virtue of making the coupling constant 
dependence of the CPO correlation functions restricted 
mostly to the overall factors.
This feature of our formulation
enables us to test the prediction (\ref{ratio_gravity})
of the AdS/CFT correspondence already for quite a small matrix size.
Our results suggest that the relation extends to 
intermediate coupling constants as well.

In fact there is strong evidence from field theoretical analysis
in the 4d $\mathcal{N}=4$ SYM that
the non-renormalization theorem holds for
two-point and three-point correlation 
functions of CPOs \cite{D'Hoker:1998tz,Eden:1999gh}\footnote{For
CPOs with $\Delta=2$, in particular, there is an independent argument for 
the non-renormalization \cite{delta2,D'Hoker:1998tz,Lee:1998bxa}.},
which implies $c_{\Delta}=c_{\Delta_{1}\Delta_{2}\Delta_{3}}=1$.
It is therefore expected that the data points in fig.\ \ref{fig:ch23}
will approach $c_2=c_{222}=1$ as we take the limit (\ref{limit})
for fixed $\lambda_{\rm SYM}$, which needs to be checked.

The analysis of 
four-point functions would 
be more interesting \cite{full}
since there is evidence for the non-renormalization theorem 
only in the extremal and next-to-extremal cases \cite{Eden:2000gg},
and in fact
the AdS/CFT predicts its violation 
for the other general cases 
in the strong coupling limit \cite{Arutyunov:2000py}.
We consider it very interesting that such nonperturbative
issues in 4d $\mathcal{N}=4$ SYM
have become 
accessible 
by computer simulations.


\acknowledgments
We thank H.\ Kawai and Y.\ Kitazawa for valuable discussions.
The work of M.\ H.\ is supported by 
Japan Society for the Promotion of Science (JSPS).
The work of G.\ I.\ and S.\ -W.\ K.\ is supported by 
the National Research Foundation of Korea (KNRF) grant funded by the Korean government (MEST) (2005-0049409 and [NRF-2009-352-C00015] ).
The work of J.\ N.\ and A.\ T.\
is supported
by Grant-in-Aid for Scientific
Research (No.\ 19340066, 20540286 and 19540294)
from JSPS.


\end{document}